# Junk News on Military Affairs and National Security: Social Media Disinformation Campaigns Against US Military Personnel and Veterans

John D. Gallacher
Oxford University
john.gallacher@cybersecurity.ox.ac.uk
@john_gallacher1

Vlad Barash
Graphika
vlad.barash@graphika.com

Philip N. Howard
Oxford University
philip.howard@oii.ox.ac.uk
@pnhoward

John Kelly
Graphika
john.kelly@graphika.com

**ABSTRACT**

*Social media provides political news and information for both active duty military personnel and veterans. We analyze the subgroups of Twitter and Facebook users who spend time consuming junk news from websites that target US military personnel and veterans with conspiracy theories, misinformation, and other forms of junk news about military affairs and national security issues. (1) Over Twitter we find that there are significant and persistent interactions between current and former military personnel and a broad network of extremist, Russia-focused, and international conspiracy subgroups. (2) Over Facebook, we find significant and persistent interactions between public pages for military and veterans and subgroups dedicated to political conspiracy, and both sides of the political spectrum. (3) Over Facebook, the users who are most interested in conspiracy theories and the political right seem to be distributing the most junk news, whereas users who are either in the military or are veterans are among the most sophisticated news consumers, and share very little junk news through the network.*

**US MILITARY PERSONNEL, VETERANS AND SOCIAL MEDIA**

Social media are an important means of communication for both active duty military personnel and veterans. When on assignment, platforms like Facebook and Twitter can allow service personnel to stay in touch with family and friends back home. After service, social media allow soldiers and support staff to stay in touch with their colleagues and friends from their period of service, which performs an important role in veteran transition into civilian life.[1]

The pubic tends to place trust in military personnel and veterans,[2] making them potentially influential voters and community leaders. Given this trust and their role in ensuring national security, these individuals have the potential to become particular targets for influence operations and information campaigns conducted on social media. There are already reports of US service personnel being confronted by foreign intelligence agencies while posted abroad, with details of their personal lives gleaned from social media.[3]

We set about mapping the influence of known sources of junk political news and information that regularly craft content for an audience of US military personnel and veterans–we call such activity Veteran Operations or "VetOps". In particular, we investigate patterns of interaction between current or former military personnel who have (i) shared junk news targeted to an audience of military personnel, (ii) engaged with users who disseminate large amounts of misinformation about national security and international affairs.

**SOCIAL NETWORK MAPPING**

Visualizing social network data is one of the most powerful ways of understanding how people pass information and associate with one another. By using selected keywords, seed accounts, and known links to content, it is possible to construct large network visualizations. The underlying networks of these visualizations can then be manipulated to find communities of accounts and clusters of association. Importantly, one can then tag these associated clusters of accounts and content with political attributes based on knowledge of account history, content type, and association metrics.

Each map of social networks constructed over Twitter and Facebook allows for insight into both social structure and flow of information. In this study, we use the Graphika visualization suite to map and tag accounts based around prominent political accounts, topics, political affiliations, and geographical areas. Successfully mapping social networks also allows us to catalogue users and content and generate both descriptive statistics and statistical models that explain changes in network structure.

Social network maps are composed of "nodes" which represent the social media accounts in question. Each node is connected to one or more further nodes in the map via social relationships; digital interactions between accounts. These maps then represent patterns of connections between nodes via a Fruchterman-Reingold visualization algorithm.[4] This visualization algorithm works to place nodes in a map onto a canvas according to two principles: first, a "centrifugal force" acts upon each node to push it to the edge of the canvas; second, a "cohesive force" acts upon every connected pair of nodes to pull them closer together.

Each node in these networks belongs to a group with a shared pattern of interests, such as a collection of Facebook accounts that all like US pro-Donald Trump political pages, for example. A group is a collection of segments that are geographically, culturally, or socially similar. For example, a segment of US Trump supporters, US Libertarians, and US Constitutional Conservatives could be grouped into a "US Conservatives" group. The method of segmenting users, coding groups, and generating broad observations about association is an iterative process that involves qualitative, quantitative and computational methods. It was run on several occasions over a period of time to identify the stable and consistent communities.

A clustering algorithm automatically generates segments and groups from the sampled data. This involves first building a bipartite graph between nodes in the map and the rest of the social medium in question. This bipartite graph provides a structural similarity metric between nodes in the map, and this metric is used in combination with a clustering algorithm in order to segment a map into distinct communities. In this case hierarchical agglomerative clustering was used (see online supplement). Additional information on the grounded typology of junk news, developed over the course of studying five elections in 2016-17 is in our series of Data Memos 2016.1-2017.8.[5–8]

Over time, we found that different social media have different attributes that are effective for identifying temporally stable communities, i.e. ones that persist over time. For example, clustering Twitter users by following and follower relationships yields much more stable communities than clustering the same by mention or retweet relationship. In Facebook clustering by the "like" relationship yields similarly stable results. Therefore, within a Facebook map, all pages liked by the public pages in question are identified in the map, and the extent to which two Facebook pages like similar pages in all of Facebook generates a higher similarity metric.

The outputs of this algorithm have been extensively tested in studying social media maps of Iran and Russia.[9,10] Subsequent to clustering, the map-making process then uses supervised machine learning to generate labels for segments and groups from a set of human-labeled examples. The machine-generated labels are then manually verified by human coders.

**SAMPLING AND METHOD**

For this study, three junk news websites specializing in content on military affairs and national security issues for US military personnel and veterans were used; veteranstoday.com, veteransnewsnow.com, and southfront.org. All three of these websites are reported to show links with Russian-origin content. In late 2013 Veterans Today began publishing content from the government-charted Russian Academy of Sciences geopolitical journal New Eastern Outlook. At a similar time, its sister site, Veterans News Now, began publishing content from the Moscow think tank Strategic Culture Foundation. Similarly, the website South Front, was registered in Moscow in early 2015 and partnered with Veterans Today later that year.[11]

We use the term "junk news" to include various forms of propaganda and ideologically extreme, hyper-partisan, or conspiratorial political news and information. Much of this content is deliberately produced false reporting. It seeks to persuade readers about the moral virtues or failings of organizations, causes or people and presents commentary as a news product. This content is produced by organizations that do not employ professional journalists, and the content uses attention grabbing techniques, lots of pictures, moving images, excessive capitalization, ad hominem attacks, emotionally charged words and pictures, unsafe generalizations and other logical fallacies.

Associated social media accounts and URLs for these sites were identified and then used to map out the wider network. This network is comprised of all of the accounts to which a particular campaign is visible, including accounts which are not actively participating in the conversation but rather simply consuming information.

For the Twitter analysis, we used our seed list to identify a broad network of Twitter users that were following, mentioning, or citing content related to veterans. For the analysis of Facebook public pages, we conducted a snowball sample of public pages that directly liked or were liked by the seed pages.

**VETERAN OPERATIONS ON TWITTER**

Our Twitter dataset contains 28,467 Twitter users collected between April 2, 2017 and May 2, 2017. We collected data by identifying all Twitter accounts who followed and mentioned three prominent military-focused junk news accounts: @VeteransToday, @SouthFronteng, @veteransnewsnow. We then reduced this space of Twitter users to a set of well-connected accounts using a variant of k-core reduction (see online supplement).[12] This reduced account set contained 12,413 Twitter users. Finally, we collected all Twitter users followed by any account in the reduced account set, in order to segment this set into communities of interest.

We used a combination of Twitter's Public and Streaming APIs and the GNIP API to collect publicly available data for analyses. Twitter's Public API provides data on a) who follows whom on Twitter (100% of all data) and b) recent tweets for each user (up to 3,200 tweets by user in reverse chronological order). The Streaming API allows for constraining queries to users who use particular keywords in their tweets or users who post tweets from a specified geographical area.

Twitter limits API access in several ways: a) by limiting streaming queries to tracking a certain number of Twitter accounts, keywords, or geographical areas, b) by constraining Decahose GNIP queries to a random 10% sample of all tweets, and c) by limiting data returned from all APIs exclusively to public (not private or banned) Twitter accounts. We address limitations a and b by using a combination of public, streaming, and GNIP API queries. In other words, we perform an initial combination of GNIP and streaming API queries that generate results we can use for a more expansive public API query on tweet histories and follow relationships. We do not believe that this limitation is a concern in this study given that 88% of Twitter users have public accounts.[13]

We were able to group the 12,413 user accounts that were sampled into eight categories of affiliation. The categories emerged through network association and interpretation of the kinds of content these users distribute and indicate as "favorite". Table 1 identifies the main groupings of the audience of military junk

**Table 1: Size, Coverage and Consistency of VetOps Audience Groups on Twitter**

|  | Users N | Users % | Coverage | Consistency |
|---|---|---|---|---|
| Conservative Politics | 1,637 | 13 | 88 | 31 |
| Euro-Right | 398 | 3 | 88 | 9 |
| Government and Public Policy | 1,168 | 9 | 61 | 1 |
| International Conspiracy Theory | 1,364 | 11 | 94 | 17 |
| Liberal Politics | 840 | 7 | 92 | 10 |
| Other | 3,355 | 27 | 100 | 20 |
| Russia Focused | 1,545 | 12 | 82 | 11 |
| Veterans & Military | 2,106 | 17 | 55 | 1 |
| Total | 12,413 | | | |

**Table 2: Heterophily Index for VetOps Audience Groups on Twitter**

| Group | Conservative Politics | Euro-Right | Government and Public Policy | International Conspiracy | Liberal Politics | Other | Russia Focused | Veterans and Military |
|---|---|---|---|---|---|---|---|---|
| Conservative Politics | 5 | 4 | 3 | 2 | 2 | 3 | 2 | 3 |
| Euro-Right | - | 5 | 1 | 4 | 3 | 3 | 3 | 1 |
| Government and Public Policy | - | - | 4 | 1 | 3 | 3 | 0 | 4 |
| International Conspiracy | - | - | - | 4 | 4 | 3 | 4 | 0 |
| Liberal Politics | - | - | - | - | 5 | 3 | 3 | 2 |
| Other | - | - | - | - | - | 3 | 3 | 3 |
| Russia Focused | - | - | - | - | - | - | 5 | 0 |
| Veterans and Military | - | - | - | - | - | - | - | 5 |

**Figure 1: VetOps Audience Groups on Twitter**

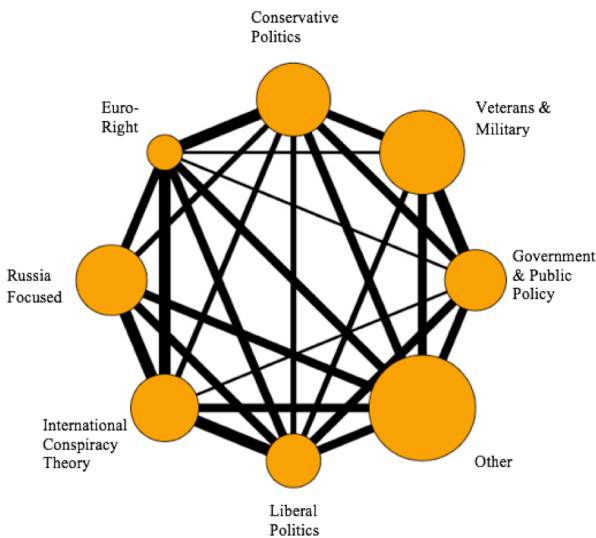

*Source: Authors' calculations from data sampled 02/4-02/5/ 2017. Note: Groups are determined through network association and our interpretation of the kinds of content these users distribute. This is a basic visualization, see comprop.oii.ox.ac.uk for a full visualization. Here, each group is represented by a single node. Node size shows the number of users in this group, while edge width shows the strength of the connection between the groups.*

news, as labeled by our iterative machine-learning and expert review.

For the entire network consuming information from military-oriented junk news websites, we can identify the number of accounts that directly share information of interest. To assist in the evaluation of this network we have computed the coverage and consistency of each group. *Coverage* refers to the percentage of all known junk news domains that a group posted links to. *Consistency* refers to the proportion of all hits on every domain that came from the group. A high value of coverage shows that the group is consuming a wide range of junk news on military affairs and national security, while a high value for consistency shows that the group is playing a large role in the distribution of such junk news. Coverage and Consistency were both calculated from a matrix of hits from groups to known Junk News, State Sponsored and Vet Ops domains.

Additionally, a value of heterophily for each combination of group pairings was calculated. This is a measure of the connections between groups in a map, whereby a ratio is calculated of the actual ties between two groups compared to the expected ties between the groups, if all the accounts in the map were evenly distributed. A natural log of the ratios is then taken along with a zero correction to create a balanced index. A higher heterophily score between groups indicates more connections between two groups, while a high heterophily score for a group to itself indicates a high number of within-group connections. It is important to note however that these scores indicate only first order connections between groups, and not second, third, or higher order connections. These values are shown in Table 2.

Figure 1 shows a basic visualization of the network map organized into eight groups. For the full visualization of all accounts separated into 45 segments within the eight groups, please see the online addendum at comprop.oii.ox.ac.uk. On the left of Figure 1 is the Russia-Focused group, which consists mostly of Pro-Putin trolls with some more internationally focused clusters such as Pro-Assad, Pro-Russia or Pro-Trump. While some clusters in this group, such as Pro-Putin Trolls, include accounts that tweet in Russian, other clusters, such as Pro-Putin Russian Trolls Abroad, tweet in a mixture of English and Russian, and can connect with English-speaking audiences. These clusters generally tweet in support of Putin's agenda, whether within the borders of Russia, in the Middle East, or as regards the US and President Trump. For the twitter network we automatically label clusters using a supervised learning algorithm. A human subject matter expert reviews the labels to ensure accuracy.

Next to the Russia-Focus group is an International Conspiracy Theory and Issue-Specific group. This group includes clusters such as Russia Today and WikiLeaks (users who follow and tweet links from both RT.com, a Russian news site, and WikiLeaks); Anti-NWO (conspiracy theorists who oppose an international "New World Order"); Pro-Palestine, and US Libertarian accounts. The unifying theme of this group is international with a conspiracy theory focus. For example, accounts in this group oppose big government, and spread conspiratorial messages about the Rothschild family.

**Table 3: Size, Coverage and Consistency of VetOps Audience Sub-Groups on Facebook**

|  | Users N | Users % | Coverage | Consistency |
|---|---|---|---|---|
| Conspiracy | 442 | 4 | 80 | 13 |
| Mental Health | 146 | 1 | 4 | 0 |
| Other | 5,032 | 45 | 96 | 33 |
| Political Left | 1,533 | 14 | 90 | 22 |
| Political Right | 740 | 7 | 82 | 24 |
| Sustainable Agriculture | 637 | 6 | 49 | 3 |
| US Military | 1,196 | 11 | 31 | 2 |
| US Veterans | 1,377 | 12 | 43 | 3 |
| Total | 11,103 |  |  |  |

**Table 4: Heterophily Index for VetOps Audience Sub-Groups on Facebook**

| Group | Conspiracy | Mental Health | Other | Political Left | Political Right | Sustainable Agriculture | US Military | US Veterans |
|---|---|---|---|---|---|---|---|---|
| Conspiracy | 6 | 3 | 4 | 4 | 4 | 4 | 1 | 1 |
| Mental Health | - | 8 | 3 | 1 | 1 | 2 | 2 | 3 |
| Other | - | - | 2 | 4 | 4 | 3 | 3 | 3 |
| Political Left | - | - | - | 5 | 2 | 3 | 1 | 1 |
| Political Right | - | - | - | - | 6 | 3 | 2 | 3 |
| Sustainable Agriculture | - | - | - | - | - | 6 | 0 | 1 |
| US Military | - | - | - | - | - | - | 5 | 4 |
| US Veterans | - | - | - | - | - | - | - | 5 |

**Figure 2: VetOps Audience Sub-Groups on Facebook**

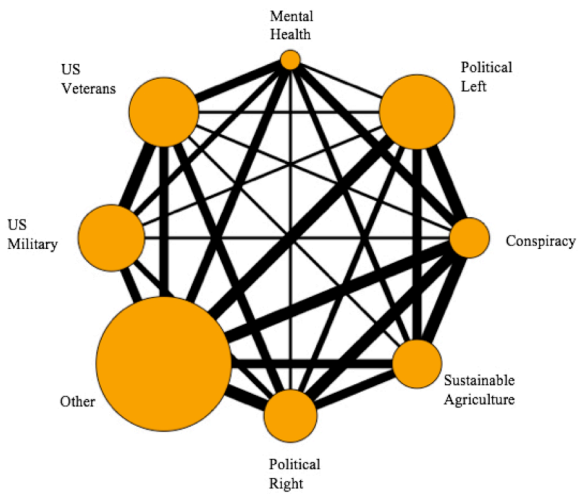

*Source: Authors' calculations from data sampled 26/5/-25/6/ 2017. Note: Sub-groups are determined through network association and our interpretation of the kinds of content these users distribute. This is a basic visualization, see comprop.oii.ox.ac.uk for a full visualization. Here, each group is represented by a single node. Node size shows the number of users in this group, while edge width shows the strength of the connection between the groups as heterophily index.*

**Table 5: Types of News Content Shared Among VetOps Users, by Sub-Groups**

|  | Junk | Professional | State Sponsored | VetOps | Total | |
|---|---|---|---|---|---|---|
|  | % | % | % | % | % | N |
| Conspiracy | 21 | 73 | 6 | 0.7 | 100 | 1,521 |
| Mental Health | 2 | 98 | 0 | 0.0 | 100 | 92 |
| Other | 16 | 81 | 3 | 0.6 | 100 | 5,287 |
| Political Left | 9 | 88 | 3 | 0.2 | 100 | 5,579 |
| Political Right | 22 | 76 | 2 | 0.2 | 100 | 3,134 |
| Sustainable Agriculture | 7 | 90 | 3 | 0.5 | 100 | 972 |
| US Military | 7 | 92 | 1 | 0.2 | 100 | 638 |
| US Veterans | 7 | 92 | 1 | 0.3 | 100 | 1,231 |
| Total | 14 | 83 | 3 | 0.4 | 100 | |

*Source: Authors' calculations from data sampled 26/5/-25/6/ 2017.*

group is partially composed of accounts that consider troll or bot accounts in their Twitter activity, and partially composed of genuine US conservatives, but the sophisticated behavior of troll and bot accounts makes precise disambiguation of these two categories difficult.

Finally, to the right of the Conservative Politics group is a Veterans and Military group. This group consists of segments devoted to the various branches of the US Military (such as Army, National Guard, or Navy and Marines), US Veterans and their support, and military families. These segments include prominent accounts for US Veteran advocacy groups, non-profits, and government organizations devoted to veteran affairs. These accounts may be troll or bot accounts in their Twitter activity, but are genuine participants in the space around US Military and Veterans.

Other, less active groups on the map include Liberal Politics, Euro-Right, and Government and Public Policy. These groups represent other contributors to the politics around US Military and Veterans affairs. It is notable that Euro-Right, which includes a UKIP cluster, is present on this map, despite the map's focus on US veterans.

The Other group includes accounts that did not meet a connectivity threshold for belonging in any one cluster. Such accounts included those labeled as Central and Eastern European Politics, Foreign Policy, as well as Social Media Marketing, Pop Culture and Spam accounts.

Overall this presents a picture with three distinct categories. First, we find genuine accounts concerning US Military and Veterans affairs. Second, we identify a mixture of genuine accounts and possible troll accounts in the space of US conservative politics. Third, we identify accounts whose activity may involve trolls or bots in the space of pro-Putin Russian politics and international conspiracy theories. The Fruchterman-Rheingold algorithm places accounts in the second category between those of the other categories, suggesting that the US conservative politics group is a mediator or network bridge that allows for a flow of information between Russian troll networks and US Military Personnel and veterans on Twitter.

When combined with the heterophily index in Table 2 these values can help clarify the observations. The highest value of heterophily between the Russia-focused

In the middle of Figure 1, at the top, is a US Conservative Politics group. This group includes a breadth of US conservative communities, from supporters of President Trump to Tea Partiers to Constitutional Conservatives and conservative pundits. This group includes accounts related to InfoWars and other news websites that have been accused of spreading fake news and conspiracy theories in the past[14]. This group and any other group is with the Conspiracy Theory group at 4 followed by the Euro-right group at 3. All other levels of heterophily are lower, suggesting that the Russian focused group, while present in the network, is fairly isolated, mostly having ties with fringe groups without having formed deep direct connections within US Military and Veteran communities online. Conversely, the level of connection between US Military and Veteran Networks and Conservative Politics at 3 is substantial enough to allow for the flow of information. An additional interesting finding is the high level of connection between US Liberal Politics and International Conspiracy Groups suggesting that this might not be a wholly conservative phenomenon.

**VETERAN OPERATIONS ON PUBLIC FACEBOOK PAGES**

Using the content that was distributed by users in the Twitter phase of this research, we proceeded to map the public Facebook pages that are sharing content from veteran misinformation campaigns.

We harvested Facebook public page seeds from the Twitter network and performed a snowball sample to discover the wider Facebook network around these key online interest groups. This was then combined with the network discovered by an initial Facebook snowball sample based on known military-oriented junk news. This snowball sampling method involved collecting all the pages that either directly liked or were liked by the Facebook accounts of the known junk news sites.

This sampling resulted in a network of 11,103 public Facebook pages. From this set we collected all posts made in the last year (8,178,004), extracted all URLs from the posts, and analyzed the pattern of web citations across the major groupings in the VetOps Facebook Network. Additionally, we collected the share counts for all posts containing these URLs in order to measure the degree to which web content from various sources is shared more widely across the Facebook network (this value includes shares that occur on private pages).

Table 3 and Figure 2 shown the groups generated from the Facebook sample, along with heterophily index and the network map. On Facebook eight groups were identified and labeled: Conspiracy Theory, Mental Health, Political Left, Political Right, Sustainable Agriculture, US Military, US Veterans, and Other. These groupings are similar to those found in the Twitter network, however with a few key differences: Initially the map shows an absence of activity outside the US. This is expected, as the map includes only US-focused clusters. Secondly, the Facebook map has an additional grouping of Sustainable Agriculture. While this may initially appear out of place in a map of US Veterans, a deeper qualitative analysis of this group revealed it to be a frontier of conspiracy theory activity, including both ostensibly right-wing and left-wing accounts. Thirdly, the Facebook map includes a Mental Health group, including communities focused on sobriety, addiction recovery and life coaching or meditation. Finally, the Facebook map includes a group on Fringe Conservative US politics including survivalists and preppers ("prepared") communities.

The heterophily index for this network indicates that the US Military and US Veteran Networks have deep connections with each other with a value of 4, while the Facebook map mirrors the Twitter map in showing developed connections between both the Political Left and Political Right with the Conspiracy Theory Group. There are also dense connections between the Conspiracy Theory Group and the Sustainable Agriculture Group with a value of 4.

The Other group includes issue segments concerning Anarchists, Animal Lovers, News and US Conservatives, and Syria. The high heterophily index for this group is found to be coming from interactions between these segments, which are not related to the rest of the map. These segments are fairly small, with the largest being News & US Conservatives.

In addition to the grouping of Facebook pages we were also able to perform an analysis of the content shared across the network in the form of URL links to external sites. We collected all URLs shared by any member of the groups identified in the network. We then classified types of news content into four categories; Junk News, Professional News, State Sponsored News, and news coming from the original VetOps accounts. This classification was based on a known dictionary of news sites, generated through manually coding the base URL of each site using a grounded typology in previous research—for detailed description see COMPROP Data Memo 2017.6.[15] This analysis is presented in Table 5, and shows that legitimate professional journalistic content was shared far more widely than junk news across the Facebook network, at a ratio of six links to professional news content for every one link to junk content. Additionally, it is shown that the Political Right group shared the highest proportion of Junk News across the Facebook network, 22%, followed by the Conspiracy Group, 21%. Conspiracy groups also shared the highest proportion of content that could be attributed to a foreign state actor with 6% of content shared in this category. Finally, both US Military and US Veteran Groups shared a low but significant proportion of Junk News content at 7% and showed a small but present interaction with VetOps content at 0.2-0.3% of total shares.

When comparing the number of times that individual content was shared across the network, we found that state sponsored and junk news content tended to have specific "amplifier" accounts. These accounts tend to do a lot of sharing, but the content they push out tends not be further shared by other users. Overall, VetOps pages and content were not shown to be especially influential on Facebook, in contrast to our results from Twitter analysis.

**CONCLUSIONS**

The social networks mapped over Twitter and Facebook include both genuine accounts created by the US military organizations, by service personnel and veterans themselves, and by groups seeking to influence those users. Some of the accounts are pro-Putin accounts pushing out significant amounts of Russian-oriented content. While Russia-aligned user accounts have built some links to an audience of current and former US military personnel, the level of this engagement is deeper on Twitter than on Facebook.

We find that on Twitter there are significant and persistent interactions between current and former military personnel and a broad network of Russia-focused accounts, conspiracy theory focused accounts, and European right-wing accounts. These interactions are often mediated by pro-Trump users and accounts that identify with far-right political movements in the US.

**ONLINE SUPPLEMENTS AND DATA SHEETS**
Please visit comprop.oii.ox.ac.uk for additional material related to the analysis, including (a) high-resolution maps of the networks for both Twitter and Facebook, showing all accounts separated into 45 segments within the 8 groups, (b) the full list of segments and groups, (c) calculation of heterophily scores (d) more detailed explanation of the hierarchical agglomerative clustering algorithm used to create groupings, (e) k-core reduction used to reduce set of Twitter users.

**ABOUT THE PROJECT**
The Project on Computational Propaganda (www.politicalbots.org) involves international, and interdisciplinary, researchers in the investigation of the impact of automated scripts—computational propaganda—on public life. *Data Memos* are designed to present quick snapshots of analysis on current events in a short format. They reflect methodological experience and considered analysis, but have not been peer-reviewed. *Working Papers* present deeper analysis and extended arguments that have been collegially reviewed and that engage with public issues. The Project's articles, book chapters and books are significant manuscripts that have been through peer review and formally published.

**ACKNOWLEDGMENTS AND DISCLOSURES**
The authors gratefully acknowledge the support of the (1) National Science Foundation, "EAGER CNS: Computational Propaganda and the Production / Detection of Bots," BIGDATA-1450193, 2014-16, Philip N. Howard, Principle Investigator; (2) the European Research Council, "Computational Propaganda: Investigating the Impact of Algorithms and Bots on Political Discourse in Europe," Proposal 648311, 2015-2020, Philip N. Howard, Principal Investigator, and (3) the Engineering and Physical Sciences Research Council (EPSRC). The project gratefully thanks the Ford Foundation for their support. Project activities were approved by the University of Washington Human Subjects Committee, approval #48103-EG and the University of Oxford's Research Ethics Committee. Any opinions, findings, and conclusions or recommendations expressed in this material are those of the authors and do not necessarily reflect the views of the National Science Foundation, the European Research Council or the Engineering and Physical Sciences Research Council or the University of Oxford.

**Junk News on Military Affairs and National Security: Social Media Disinformation Campaigns Against US Military Personnel and Veterans**

COMPROP ONLINE SUPPLEMENT TO DATA MEMO 2017.9 / 09 OCTOBER 2017

John D. Gallacher
Oxford University
john.gallacher@cybersecurity.ox.ac.uk
@john_gallacher1

Vlad Barash
Graphika
vlad.barash@graphika.com

Philip N. Howard
Oxford University
philip.howard@oii.ox.ac.uk
@pnhoward

John Kelly
Graphika
john.kelly@graphika.com

**Appendix 1. The Audience for Veterans Operations and Related Content on Twitter**

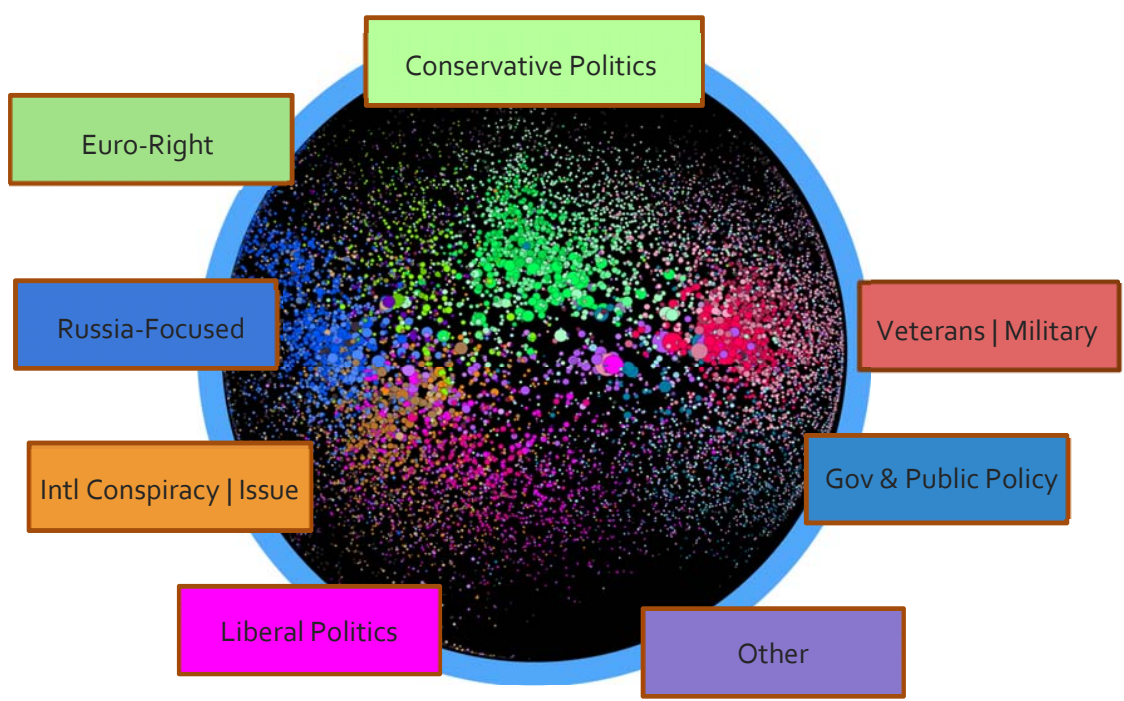

*Figure 1. Full visualisation of the audience for Veterans Operations and Related Content on Twitter
Authors' calculations from data sampled 02/4/-02/5/ 2017*

Each node in this network represents an account on Twitter. Each node belongs to both a broad group and a smaller segment within that group. The size of the node is proportional to the number of other map nodes that follow it on Twitter. The colour of the node is based on its parent segment.

A segment is a collection of nodes with a shared pattern of interest while a group is a collection of segments that are geographically, culturally, or socially similar.

The nodes are placed within the map using a Fruchterman-Reingold visualization algorithm. This works to place nodes into the map according to two principles: first, a "centrifugal force" acts upon each node to push it to the edge of the canvas; second, a "cohesive force" acts upon every connected pair of nodes to push them closer together.

**Full list of Groups and Segments for the Twitter Map**

| Group | Segment | Group | Segment | Group | Segment |
|---|---|---|---|---|---|
| Russia-Focus | Pro Putin Trolls / Pols | Conservative Politics | Conservative Pundits / Celebs | Gov & Public Policy | Public Health |
| | Foreign Policy Journos / MENA | | Trumpista | | Beltway Polit /Congress |
| | Pro Putin Russian Trolls Abroad | | True American Patriotism | | Tech and Finance News |
| | Pro Assad / Russia / Trump | | Real Donald Trump | | Nonprofit / Eco / Education |
| | Pro Putin Russians / Ukraine | | Conservative Pundit / Fox | | US Gov / Emergency Response |
| Intl Conspiracy / Issue | Anti-NWO | | Tea Party / Guns | Other | Pop Culture |
| | Pro-Palestine | | Pro-Trump Core | | Pop Culture |
| | US Libertarian | | Constitutional Conservatives | | Celeb / Wrestling Focus |
| | Intl RT and Wikileaks | Liberal Politics | Pro-Bernie / Resist | | SMM Inspiration |
| Veterans / Military | US Military 2 | | UK Left | | Central / Eastern Europe Politics |
| | US Military / Navy / Marines | | Progressives | | Foreign Policy Intl / US |
| | Defence Industry | | US Liberals | | SMM Motivation |
| | Conservative / Veteran 1 | | Prog Journo / Activism | | |
| | Army / National Guard | Euro-Right | UKIP | | |
| | Veterans | | White Identity | | |
| | Veterans | | | | |
| | Veteran Support | | | | |
| | Military Families | | | | |

**Appendix 2. Audience for Veterans Operations and Related Content on Facebook**

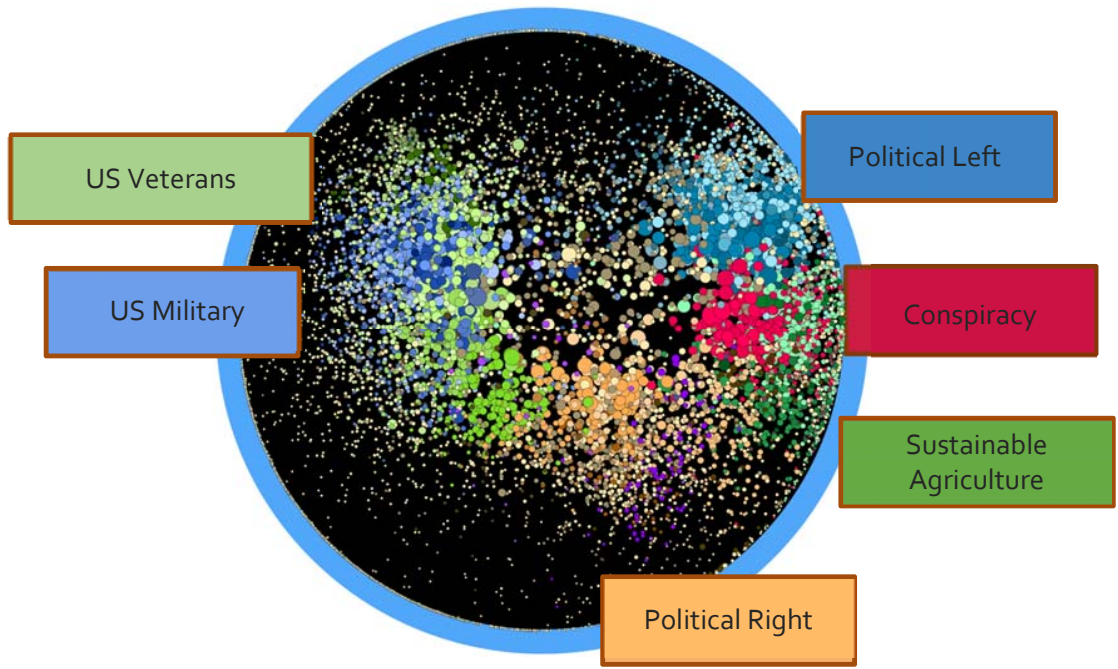

*Figure 2. Full visualisation of the audience for Veterans Operations and Related Content on Facebook*
*Authors' calculations from data sampled 26/5/-25/6/ 2017*

Each node in this network represents a public page on Facebook. The size of the node corresponds to the number of other nodes in the map that like the page on Facebook. Each node belongs to both a broad group and a smaller segment within that group. A segment is a collection of nodes with a shared pattern of interest while a group is a collection of segments that are geographically, culturally, or socially similar

Again, a Fruchterman-Reingold visualization algorithm is used to place nodes within the map.

**Full list of Groups and Segments for the Facebook Map**

| Group | Segments | Group | Segments | Group | Segments |
|---|---|---|---|---|---|
| US Veterans | Veterans Networks/ Disability | Political Right | Libertarian / Youth | Conspiracy | Conspiracy / RT |
| | Veteran Support|/ Families | | Libertarian Institutions | | Truth / Truthers |
| | Veterans Networks | | House Republicans | | Far Right / Conspiracy |
| | US Military / Veteran Support | | Conservative Media | Sustainable Agriculture | Organic / Sustainable Ag |
| | Military Gear / Weapons | | Conservative Pundits | | Health / Nutrition |
| | US VA | | Hard Conservative | | Anti-GMO |
| | VA Hospitals | | Prepper / Survivalist | | Small Farms / Canning |
| | Veterans Support | | Conservative and Pro-Israel | | Natural Living / Organic |
| | US Mil Community | | Guns | Mental Health | Mental Health |
| | American Legion | | Libertarian and End Fed Reserve | | Life Coach and Meditation |
| US Military | US Army / Armed Forces | | US Far Right and Anti-Immigrant | | Sobriety and Addiction Recovery |
| | US Military | | Conservative/ Townhall | Other | News and US Conservative |
| | US Navy | | Hard Right / Pro-Military | | Animal Lovers and Rescue |
| | US Military Europe / Africa | Political Left | Conservation | | Anarchist |
| | US Army / National Guard | | Womens Issues|Intl | | Syria and Assad |
| | US Coast Guard | | Western Liberal Media | | |
| | Navy Seals / Special Ops | | Labor Rights|Unions | | |
| | US Air Force | | Intl Occupy | | |
| | US Marines | | Progressive Dems | | |
| | National Guard | | US Occupy | | |
| | US Forces / Korea | | Intl Direct Democracy|Anon | | |
| | | | Occupy | | |
| | | | Occupy|Economic Inequality | | |

**Additional Methodological Descriptions**

**Appendix 3. Heterophily Index**

For every pairing of groups within a network map, a value of heterophily can be calculated. This is a measure of the level of connection between the groups. In order to determine this a ratio is calculated of the actual ties between two groups compared to the expected ties between the groups if all the accounts in the map were evenly distributed.

The natural log of these ratios is then taken, along with a zero correction to create a balanced index and ensure that all values are displayed in a positive form.

$$\text{Ratio of Ratios}_T = \frac{\dfrac{\text{Connections}_{\text{pairing}}}{\sum_{\text{all pairings}} \text{Connections}}}{\dfrac{\text{Connections}_{\text{pairing}}}{\sum_{\text{all pairings}} \text{Connections}}}$$

        Expression A: Ratio of Two Ratios

This heterophily index is therefore created through a ratio of two ratios. The ratio of these two ratios reveals whether two nodes have about the proportion of links they should have given its size. This is displayed in Expression A, where a pairing of groups is calculated as having a measure of connections in balance with its share of all the connections.

Half the distribution of possible values from this ratio of ratios ranges from 0 to 1 (a disproportionately small share of connections in a group given its size) and the other half ranges from 1 to +infinity (a disproportionately large share of connections in a group given its size). However, by taking the natural log of the ratio of ratios the index will become more balanced: from -infinity to 0 becomes less than proportionate share, and from 0 to +infinity becomes more than proportionate share.

For example, take a three-group network (A, B and C). If nodes in group A have a total of ten connections, and there are ten nodes in each group, then the expected connections between A and B will be 3.33. If, in reality, the nodes in group A actually have all ten connections to nodes in group B then this connection is stronger than expected. The heterophily score for groups A and B = 10/3.33 = 3.0. The natural log of this is then taken along with a zero correction across the range of heterophily values.

A greater heterophily index indicates a denser pattern of connections between the two groups. It is important to note however that these scores indicate only first order connections, not second or third order connections.

**Appendix 4. Clustering for groups and Segments**

In order to generate segments and groups for each map it is necessary to employ a clustering algorithm.

This involves first building a bipartite graph between nodes in the map and the rest of the social medium in question. This bipartite graph provides a structural similarity metric between nodes in the map.

This was then used in combination with a hierarchical agglomerative clustering algorithm in order to segment a map into distinct communities. This is a 'bottom up' approach whereby each observation starts in its own cluster, and pairs of clusters are merged as one moves up the hierarchy.

Twitter maps are clustered based on follower relationships, since mentions relationships have been shown to overemphasize the news cycle and salient external events. Facebook networks are clustered based on page likes.

**Appendix 5. K-core reduction**

To identify and map the 'discussion core' of the most active, connected, and influential users, we performed a k-core reduction to reduce the total collected set of Twitter users from the initial data collection into a set of well-connected accounts. This produces a maximally connected subgraph of active nodes with degree of connection at least *'k'*.

This degree of connection, k, can be thought of as the number of links between each node in the graph. For example, selecting a k value of 0 for the reduction not remove any nodes from the graph, since each node must have 0 connections or greater. Selecting a k value of 1 would remove all of the nodes that have no connections to other nodes in the graph. Selecting a k value of 2 would remove all nodes with fewer than 2 connections, and so on.

A value of k was selected such that the k-core consisted of 12,413 users. This value was found to be a sufficiently large group to represent the major sets of highly active users, but not so large as to make clustering and visualization impractical.